\documentclass[aps,twocolumn,10pt,longbibliography]{revtex4-1}
\usepackage{graphicx}
\usepackage{txfonts}
\usepackage[T1]{fontenc}
\usepackage{epstopdf}
\usepackage{verbatim}
\usepackage{color}
\usepackage{soul} %\st \hl
\usepackage{fancybox}

\begin{document}
\title{Topological Photonics}
\author{Ling Lu}
\email{linglu@mit.edu}
\author{John D. Joannopoulos}
\author{Marin Solja\v{c}i\'{c}}
\affiliation{Department of Physics, Massachusetts Institute of Technology, Cambridge, Massachusetts 02139, USA}
\date{\today}
\def\clr2{red}
\begin{abstract}
The application of topology, the mathematics studying conserved properties through continuous deformations, is creating new opportunities within photonics, bringing with it theoretical discoveries and a wealth of potential applications.
This field was inspired by the discovery of topological insulators, in which interfacial electrons transport without dissipation even in the presence of impurities. Similarly, the use of carefully-designed wave-vector space topologies allows the creation of interfaces that support new states of light with useful and interesting properties. In particular, it suggests the realization of unidirectional waveguides that allow light to flow around large imperfections without back-reflection. The present review explains the underlying principles and highlights how topological effects can be realized in photonic crystals, coupled resonators, metamaterials and quasicrystals.
\end{abstract}

\maketitle
\def\clri{red}
\def\clr{black}
\def\clrl{black}

Frequency, wavevector, polarization and phase are degrees of freedom that are often used to describe a photonic system.
In the last few years, topology --a property of a photonic material that characterizes the quantized global behavior of the wavefunctions on its entire dispersion band-- has been emerging as another indispensable ingredient, opening a path forward to the discovery of fundamentally new states of light and possibly revolutionary applications.
Possible practical applications of topological photonics 
 include photonic circuitry less dependent on isolators and slow light insensitive to disorder.

Topological ideas in photonics branch from exciting developments in solid-state materials, along with the discovery of new phases of matter called \emph{topological insulators}~\cite{Hasan2010TopologicalReview,Qi2011TIreview}. Topological insulators, being insulating in their bulk, conduct electricity on their surfaces without dissipation or back-scattering, even in the presence of large impurities. The first example was the \emph{integer quantum Hall effect}, discovered in 1980. In quantum Hall states, two-dimensional~(2D) electrons in a uniform magnetic field form quantized cyclotron orbits of discrete eigenvalues called Landau levels. When the electron energy sits within the energy gap between the Landau levels, the measured edge conductance remains constant within the accuracy of about one part in a billion, regardless of sample details like size, composition and impurity levels. In 1988, Haldane proposed a theoretical model to achieve the same phenomenon but in a periodic system without Landau levels~\cite{haldane1988model}, the so-called \emph{quantum anomalous Hall effect}.

Posted on arXiv in 2005, Haldane and Raghu transcribed the key feature of this electronic model into photonics~\cite{Haldane:2008-PRL,raghu2008analogs}. They theoretically proposed the photonic analogue of the quantum (anomalous) Hall effect in photonic crystals~\cite{Joannop:book}, the periodic variation of optical materials, molding photons the same way as solids modulating electrons. Three years later, the idea was confirmed by Wang et al., who provided realistic material designs~\cite{Wang2008PRL} and experimental observations \cite{Wang2009}. Those studies spurred numerous subsequent theoretical~\cite{hafezi2011robust,fang2012realizing,khanikaev2013photonic,lu2013weyl,skirlo2014multimode} and experimental investigations~\cite{kraus2012topological,rechtsman2013photonic,hafezi2013imaging}.

Back-reflection, the source of unwanted feedback and loss, in ordinary waveguides is a main obstacle to large scale optical integration.
The works cited above demonstrated unidirectional edge waveguides transmit electromagnetic waves without back-reflection even \emph{in the presence of arbitrarily large disorder}: this ideal transport property is unprecedented in photonics. Topological photonics promises to offer unique, robust designs and new device functionalities to photonic systems by providing immunity to performance degradation induced by fabrication imperfections or environmental changes.

In this review, we present the key concepts, experiments, and proposals in the field of topological photonics.
Starting with an introduction to the relevant topological concepts, we introduce the 2D quantum Hall phase through the stability of Dirac cones~\cite{Haldane:2008-PRL,raghu2008analogs}, followed by its realizations in gyromagnetic photonic crystals~\cite{Wang2008PRL,Wang2009,skirlo2014multimode}, in coupled resonators~\cite{hafezi2011robust,fang2012realizing,hafezi2013imaging} and waveguides~\cite{rechtsman2013photonic}, in bianisotropic metamaterials~\cite{khanikaev2013photonic} and in quasicrystals~\cite{kraus2012topological}. We then extend our discussions to three dimensions, wherein we describe the stability of line nodes and Weyl points and their associated surface states~\cite{lu2013weyl}. We conclude by considering the outlook for further theoretical and technological advances.

\section{Topological phase transition}
\begin{figure}[ht]
\includegraphics[width=0.5\textwidth]{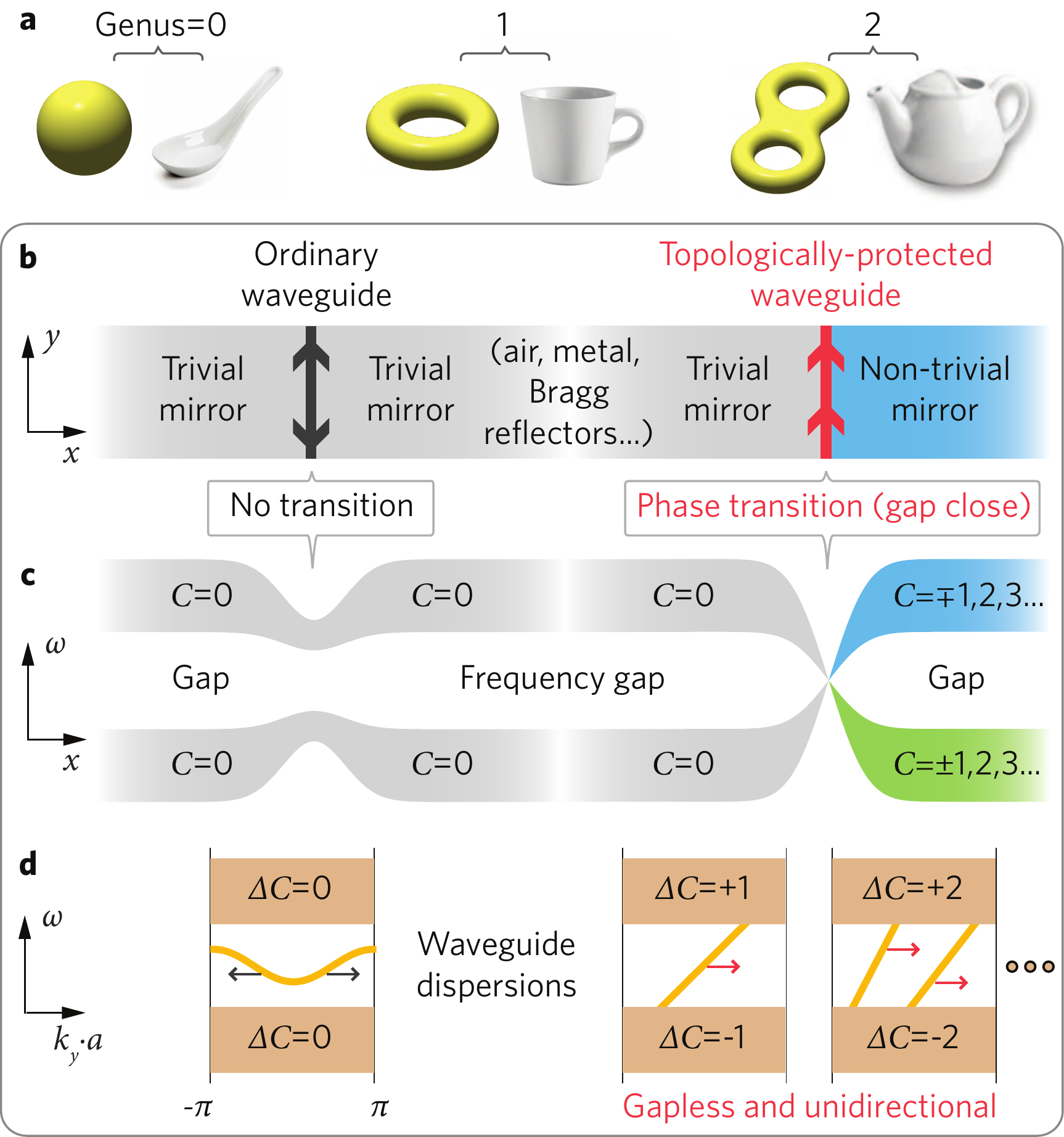}
\caption{
\textbf{Topological phase transition.}
\textbf{a}, Six objects of different geometries can be grouped into three pairs of topologies. Each pair has the same topological invariant called genus. 
\textbf{b}, Illustration of two waveguides formed by mirrors of different (right) and same (left) topologies.
\textbf{c}, Frequency bands of different topologies cannot transit into each other without closing the frequency gap. A topological phase transition takes place on the right, but not on the left.
\textbf{d}, Interfacial states have different connectivity with the bulk bands, depending on the band topologies of the bulk mirrors. $a$ is the period of the waveguide propagating along $y$. $\Delta C$ is the change in Chern number between the corresponding bulk bands on the right and left of the waveguide. The magnitude of $\Delta C$ equals the number of gapless interfacial modes and the sign of $\Delta C$ indicates the direction of propagations.
}
\label{Fig:0_intro}
\end{figure}

\emph{Topology} is the branch of mathematics that concerns quantities that are preserved under continuous deformations. For example, the six objects in Fig. \ref{Fig:0_intro}a all have different geometries; but there are only three different topologies. The yellow sphere can be continuously deformed into the white spoon, so they are \emph{topologically equivalent}. The torus and coffee cup are also topologically equivalent, and so too are the double torus and tea pot. Different topologies can be mathematically characterized by integers called \emph{topological invariants}, quantities that remain the same under arbitrary continuous deformations of the system. For the above closed surfaces, the topological invariant is the genus, and it corresponds to the number of holes within a closed surface. Objects having the same topological invariant are topologically equivalent: they are in the same \emph{topological phase}. Only when a hole is created or removed in the object does the topological invariant change. This process is a \emph{topological phase transition}.

Material-systems in photonics have topologies, defined on the dispersion bands in the reciprocal (wavevector) space. The topological invariant of a 2D dispersion band is the \emph{Chern number} ($C$ in Box 1), a quantity that characterizes the quantized collective behavior of the wavefunctions on the band. Once a physical observable can be written as a topological invariant, it only changes discretely; thus, it will not respond to continuous small perturbations. These perturbations can be arbitrary continuous changes in the material parameters.

Optical mirrors reflect light of a given frequency range: light reflects due to the lack of available optical states inside the mirror. Mirrors, that is, have \emph{frequency gaps} in analogy to the energy gaps of insulators.
The sum of the Chern numbers of the dispersion bands below the frequency gap labels the topology of a mirror. This can be understood as the total number of ``twists'' and ``un-twists'' of the system up to the gap frequency. 
Ordinary mirrors like air (total internal reflection), metal, or Bragg reflectors all have zero Chern numbers- they are \emph{topologically trivial}. Mirrors with non-zero Chern numbers are instead \emph{topologically non-trivial}. 

The most fascinating and peculiar phenomena take place at the interface where two mirrors having different topological invariants join together. 
The edge waveguide formed by these two topologically in-equivalent mirrors  (right of Fig. \ref{Fig:0_intro}b) is topologically distinct from an ordinary waveguide, which is formed between topologically equivalent mirrors (left of Fig. \ref{Fig:0_intro}b). The distinction lies in the frequency spectra of their edge modes inside the bulk frequency gap. On the left of Fig. \ref{Fig:0_intro}c, the two frequency bands both have zero Chern numbers, so they can directly connect across the interface without closing the frequency gap. However, when the two mirrors have different Chern numbers, topology does not allow them to connect to each other directly. A topological phase transition must take place at the interface: this requires closing the frequency gap, neutralizing the Chern numbers, and reopening the gap. This phase transition, illustrated on the right of Fig. \ref{Fig:0_intro}c, ensures \emph{gapless} frequency states at the interface: there must exist edge states at all frequencies within the gap of the bulk mirrors. The gapless spectra of the edge states are \emph{topologically protected}, meaning their existence is guaranteed by the difference of the topologies of the bulk materials on the two sides.
In general, the number of gapless edge modes equals the difference of the bulk topological invariants across the interface. This is known as the \emph{bulk-edge correspondence}.

The topological protection of edge waveguides can also be understood in the reciprocal space. Figure \ref{Fig:0_intro}d shows the dispersion diagrams of both ordinary (left) and gapless (right) waveguides. On the left, the ordinary waveguide dispersion is disconnected from the bulk bands and can be continuously moved out of the frequency gap, into the bulk bands. On the right, however, the gapless waveguide dispersion connects the bulk frequency bands above and below the frequency gap. It cannot be moved out of the gap by changing the edge terminations.
Similar comparisons between the edge band diagrams are shown in Fig. \ref{Fig:1_Dirac}.
The only way to alter these connectivities is through a topological phase transition, i.e. closing and reopening the bulk frequency gap.

The unidirectionality of the protected waveguide modes can be seen from the slopes (group velocities) of the waveguide dispersions. 
The ordinary waveguide (Fig. \ref{Fig:0_intro}d left) supports bi-directional modes; it back-scatters at imperfections. By contrast, the topologically-protected gapless waveguides (Fig. \ref{Fig:0_intro}d right)  are unidirectional, having only positive~(negative) group velocities. In addition, there are no counter-propagating modes at the same frequencies as the one-way edge modes. This enables light to flow around imperfections with perfect transmission; except going forward, light has no other options. The operation bandwidth of these one-way waveguides are as large as the size of the bulk frequency gap.

\section{From Dirac cones to quantum Hall topological phase}
\begin{figure}[ht]
\includegraphics[width=0.46\textwidth]{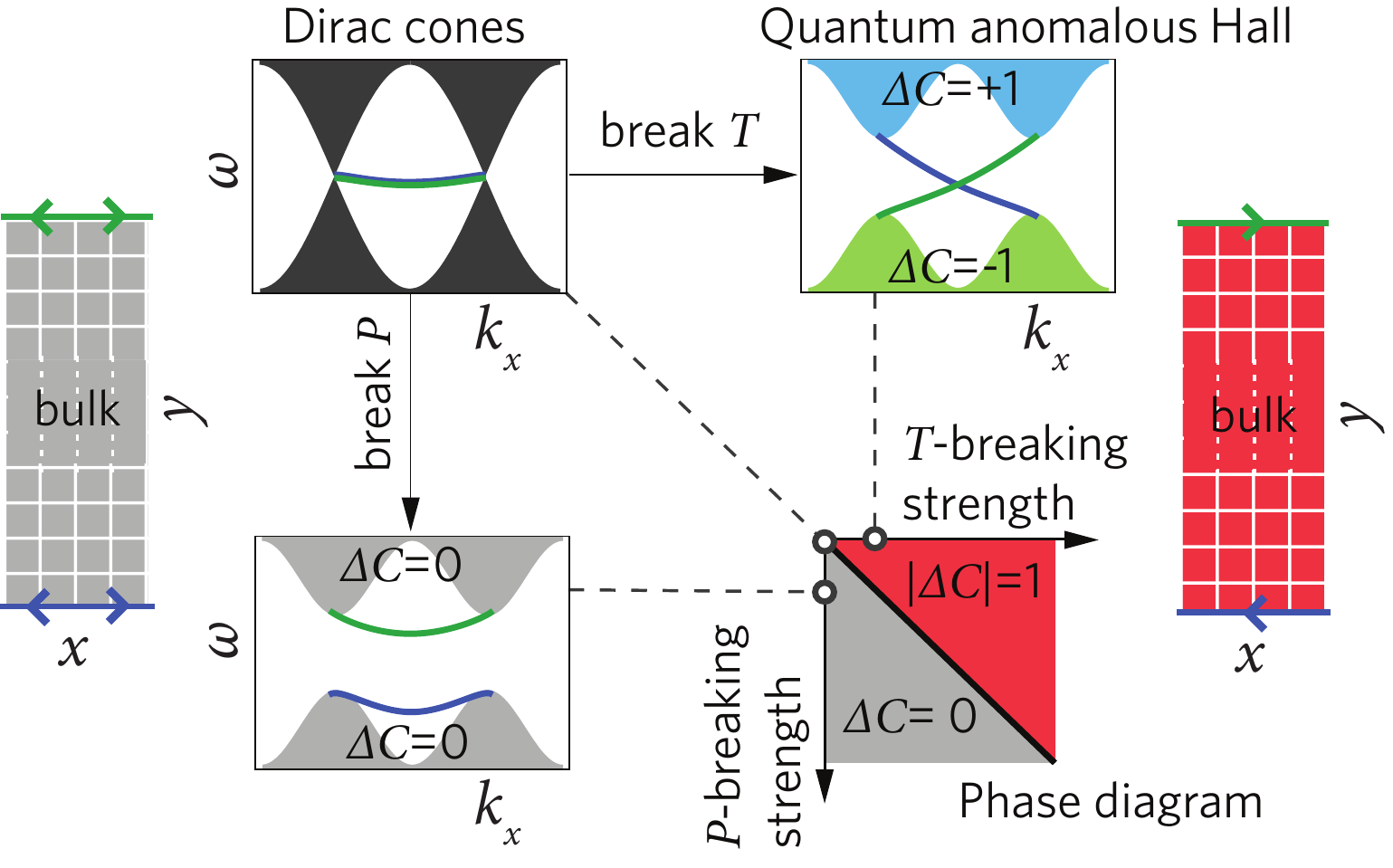}
\caption{
\textbf{Topological phase diagram of the 2D quantum Hall phase.}
A band diagram of edge states are illustrated on the top left when the bulk dispersions form a pair of Dirac cones (gray) protected by $PT$ symmetry. The green and blue colors represent edge dispersions on the top and bottom edges. When either $P$ or $T$ are broken, a bandgap can form in the bulk but not necessarily on the edges. When $T$-breaking is dominant, the two bulk bands split from one pair of Dirac degeneracies and acquire Chern numbers of $\pm1$. So there exists one gapless edge dispersion on each of the top and bottom interfaces, assuming the bulk is interfaced with topologically trivial mirrors. This $T$-breaking phase of non-zero Chern numbers are the quantum Hall phase, plotted in red in the phase diagram. 
}
\label{Fig:1_Dirac}
\end{figure}
An effective approach to find non-trivial mirrors (frequency gaps with non zero Chern numbers) is to identify the phase transition boundaries of the system in the topological phase diagram, where the bulk frequency spectrum is gapless. Then, a correct tuning of the system parameters will immediately open gaps belonging to different topological phases.
In 2D periodic systems, these phase boundaries are point-degeneracies in the bandstructure. The most fundamental 2D point degeneracy is a pair of Dirac cones of linear dispersions between two bands. In 3D, the degeneracies are line nodes and Weyl points that we will discuss later in this review.

Dirac cones are protected, in the whole 2D Brillouin zone, by $PT$ symmetry, the product of time-reversal symmetry ($T$, in Box 2) and parity ($P$) inversion. Every Dirac cone has a quantized Berry phase~(Box 1) of $\pi$ looped around it~\cite{sepkhanov2008proposed,sepkhanov2009extinction}. Protected Dirac cones generate and annihilate in pairs~\cite{Peleg2007Conical,bahat2010klein,rechtsman2013strain,schomerus2013parity,rechtsman2013topological}.  The effective Hamiltonian close to a Dirac point, in the x-y plane, can be expressed by $H(\mathbf{k})=v_xk_x\sigma_x+v_yk_y\sigma_z$, where $v_i$ are the group velocities and $\sigma_i$ are the Pauli matrices. Diagonalization leads to the solution $\omega(\mathbf{k})=\pm\sqrt{{v_x}^2{k_x}^2+{v_y}^2{k_y}^2}$. Both $P$ and $T$ map the Hamiltonian from $k$ to $-k$, but they differ by a complex conjugation:
so $(PT)H(\mathbf{k})(PT)^{-1}=H(\mathbf{k})^*$.
$PT$ symmetry requires the Hamiltonian to be real and absent of $\sigma_y=\left(\begin{array}{cc}0&-i\\i&0\end{array}\right)$ that is imaginary.
A 2D Dirac point-degeneracy can be lifted by any perturbation that is proportional to $\sigma_{y}$ in the Hamiltonian or, equivalently, by any perturbation that breaks $PT$. Therefore, breaking either $P$ or $T$ will open a bandgap between the two bands.

However, the bandgaps opened by breaking $P$~\cite{onoda2004hall} and $T$ individually are topologically inequivalent~\cite{raghu2008analogs,ochiai2009photonic}, meaning that the bulk bands in these two cases carry different Chern numbers.
The Chern number is the integration of the Berry curvature~[$\mathcal{F}(\mathbf{k})$ in Table B1] on a closed surface in the wavevector space. $\mathcal{F}(\mathbf{k})$ is a pseudovector that is odd under $T$ but even under $P$.
In the presence of both $P$ and $T$, $\mathcal{F}(\mathbf{k})=0$.
When one of $P$ and $T$ is broken, the Dirac cones open and each degeneracy-lifting contributes a Berry flux of magnitude $\pi$ to each of the bulk bands.
In the presence of $T$~($P$ broken), $\mathcal{F}(\mathbf{k})=-\mathcal{F}(\mathbf{-k})$. The Berry flux contributed by one pair of Dirac points at $\mathbf{k}$ and $\mathbf{-k}$ are of opposite signs. The integration over the whole 2D Brillouin zone always vanishes, so do the Chern numbers.
In contrast, in the presence of $P$~($T$ broken), $\mathcal{F}(\mathbf{k})=\mathcal{F}(\mathbf{-k})$. The total Berry flux adds up to $2\pi$ and the Chern number equals one. More pairs of Dirac cones can lead to higher Chern numbers~\cite{skirlo2014multimode}.
This topologically non-trivial 2D phase is colored with red in the phase diagram in Fig. \ref{Fig:1_Dirac}.
\\

\section{Gyromagnetic photonic crystals}
The first experiments~\cite{Wang2009} to realize the photonic analogue of the  quantum Hall effect were by Wang et al. at microwave frequencies. The experiment used gyromagnetic materials and introduced a uniform magnetic field to break $T$. The result is a single topologically-protected edge waveguide mode that propagates around arbitrary disorder without reflection.

These single-mode one-way waveguides can also be realized in coupled defect cavities~\cite{fang2011microscopic}, self-guide~\cite{poo2011experimental} in free-standing slabs~\cite{liu2012one} and have robust local density of states~\cite{asatryan2013local}.
They have enabled novel device designs for tunable delays and phase shifts with unity transmission~\cite{Wang2008PRL}, reflectionless waveguide bends and splitters~\cite{he2010tunable}, signal switches~\cite{zang2011edge}, directional filters~\cite{fu2010robust,fu2011unidirectional}, broadband circulators~\cite{qiu2011broadband} and slow-light waveguides~\cite{yang2013experimental}. Very recently, multi-mode one-way waveguides of large bulk Chern numbers ($|C|=2,3,4$) have been constructed by opening gaps of multiple point degeneracies simultaneously~\cite{skirlo2014multimode}, providing even richer possibilities in device functionalities.

\begin{figure}[h]
\includegraphics[width=0.5\textwidth]{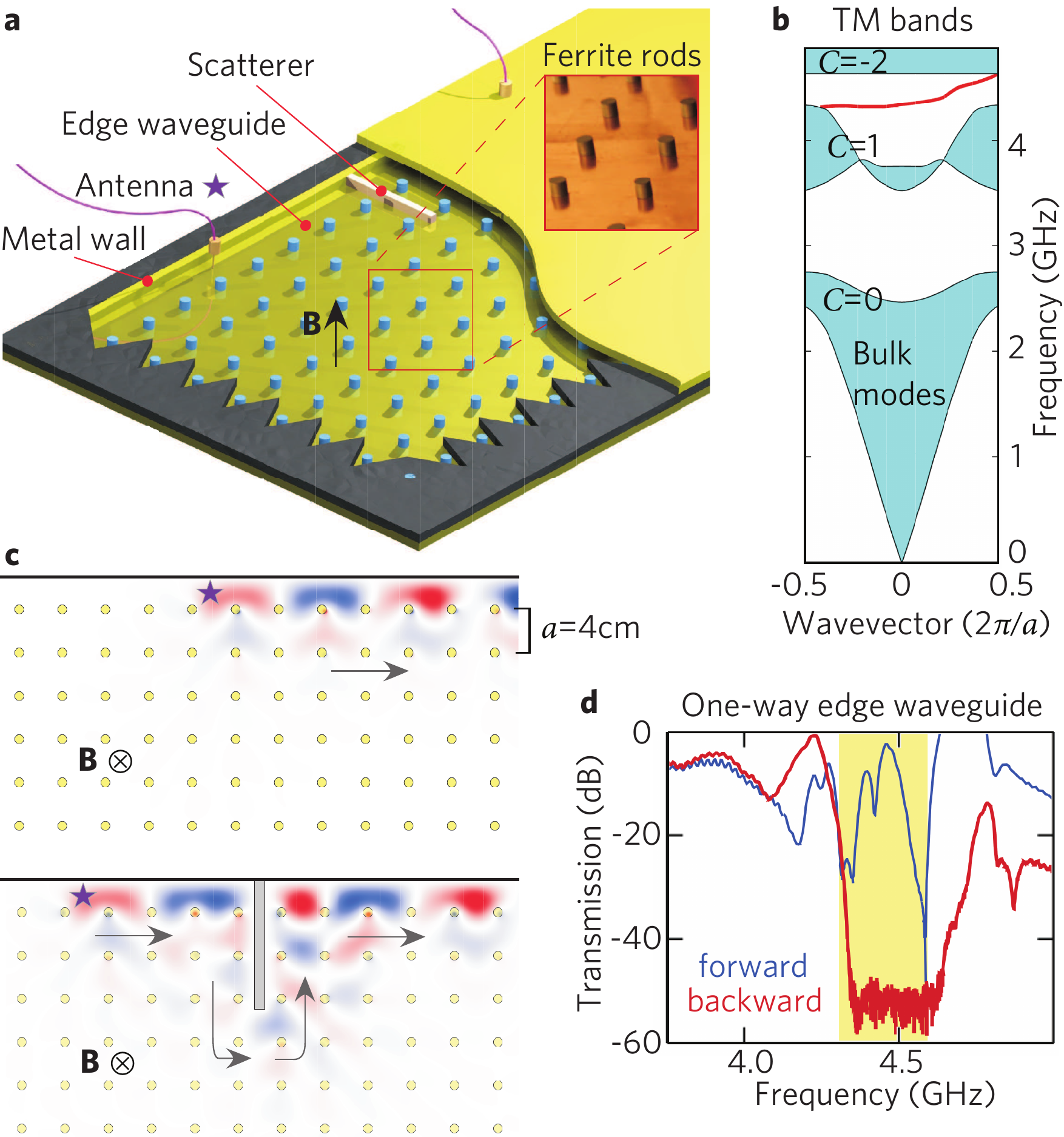}
\caption{\textbf{First experimental demonstration of the topologically-protected one-way edge waveguide at microwave frequencies.}}
\textbf{a}, Schematic of the experimental setup for measuring the one-way edge state between the metal wall and the gyromagnetic photonic crystal confined between the metallic plates to mimic the 2D TM modes. The inset is a picture of the ferrite rods that constitute the photonic crystal of lattice period $a=4cm$.
\textbf{b}, The bandstructure of the one-way gapless edge state between the second and third bands of non-zero Chern numbers.
\textbf{c}, Simulated field propagation of the one-way mode and its topological protection against large obstacles.
\textbf{d}, The measured high one-way transmission data of the edge waveguide.
\label{Fig:2_Nature-exp}
\end{figure}

The experiments in Ref.~\cite{Wang2009} were based on a 2D square lattice photonic crystal composed of an array of gyromagnetic ferrite rods confined vertically between two metallic plates to mimic the 2D transverse magnetic~(TM) modes. Shown in Fig. \ref{Fig:2_Nature-exp}a, a metal wall was added to the surrounding edges to prevent radiation loss into air. Without the external magnetic field, the second and third TM bands are connected by a quadratic point-degeneracy composed of a pair of Dirac cones~\cite{Chong2008quadratic}.
Under a uniform static magnetic field (0.2 Tesla) that breaks $T$, anti-symmetric imaginary off-diagonal terms develop in the magnetic permeability tensor ($\mu$). The quadratic degeneracy breaks and a complete bandgap forms between the second and third bands, both having non-zero Chern numbers. The red dispersion line in Fig. \ref{Fig:2_Nature-exp}b is the gapless edge state inside the second bandgap having only positive group velocities at around 4.5~GHz. Numerical simulation results in the top of plot Fig. \ref{Fig:2_Nature-exp}c verified that an antenna inside the waveguide can only emit into the forward direction in the bulk frequency gap. The experimental transmission data in Fig. \ref{Fig:2_Nature-exp}d shows that the backward reflection is more than five orders of magnitude smaller than the forward transmission after propagating over only eight lattice periods.
More importantly, there is no increase in the reflection amplitude even after inserting large metallic obstacles in the experiments, as illustrated in the lower plot of Fig. \ref{Fig:2_Nature-exp}c. Indeed, new one-way edge modes automatically form wherever a new interface is created, providing a path for light to circumvent the obstacle. This is precisely the topological protection provided by the bulk of the photonic crystal containing non-zero Chern numbers.

We note there exist other types of one-way waveguides that break $T$~\cite{yu2008one}, but they are not protected by topology. In general, magnetic responses are very weak in optical materials.
Therefore, realizations at optical frequencies remain a challenge.

\section{Coupled resonators}

Photons in an array of coupled resonators are similar to electrons in an array of atoms in solids. The photon couplings between the resonators can be controlled to  form topologically non-trivial frequency gaps with robust edge states.
Researchers obtained the photonic analogues of the integer quantum Hall effect by constructing both static and time-harmonic couplings that simulate the electron's behavior in a uniform magnetic field.
When the $T$-breaking is implemented by accurate time-harmonic modulations, unidirectional edge waveguides immune to disorder can be realized at optical frequencies.

\begin{figure}[h]
\includegraphics[width=0.5\textwidth]{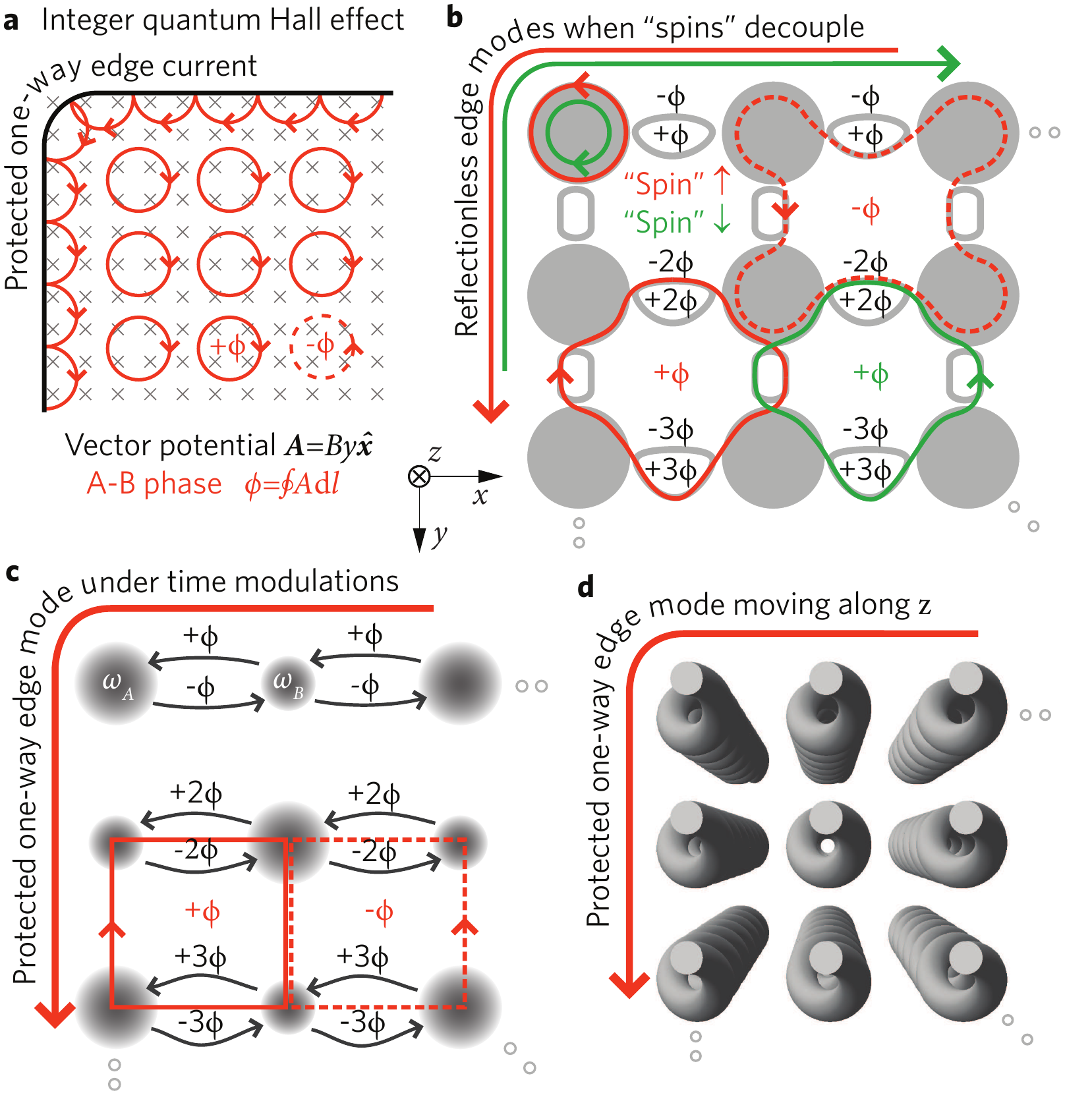}
\caption{\textbf{Quantum Hall phase of electrons in a magnetic field and of photons in coupled resonators exhibiting an effective magnetic field.}
\textbf{a}, Illustration of the cyclotron motions of electrons in a static magnetic field ($B\hat{z}$). The vector potential increases linearly in $y$.
\textbf{b}, A 2D lattice of photonic whispering-gallery resonators coupled through static waveguides. The horizontal coupling phases increase linearly in $y$.
The two ``spins'' of the whispering-gallery resonators are degenerate in the effective magnetic field.
\textbf{c}, A 2D lattice of photonic resonators consisting of two types of single-mode cavities. The nearest neighbors are coupled through time-domain modulations with horizontal phases linearly increasing in $y$; this breaks $T$.
\textbf{d}, An array of helical photonic waveguides, breaking $z$ symmetry, induces harmonic modulations on photons propagating in it.
}
\label{Fig:3_Crows}
\end{figure}

In electronic systems, the first quantum Hall effect was observed in a 2D electron gas subject to an out-of-plane magnetic field.
As illustrated in Fig. \ref{Fig:3_Crows}a, the bulk electrons undergo localized cyclotron motions, while the unidirectional edge electrons have an extended wavefunction. Again, the number of the gapless edge channels equals the Chern number of the system.
Here, the physical quantity describing the magnetic field is the vector potential, that can be written in the form $\mathbf{A}=By\hat{x}$.
An electron accumulates Aharonov-Bohm (A-B) phase of $\phi=\oint_{}\mathbf{A(r)}\cdot\mathrm{d}\mathbf{l}$ after a closed loop~(also see Table B1). An electron going against the cyclotron motion acquires $-\phi$ phase indicated by a dotted circle in Fig. \ref{Fig:3_Crows}a, so it has a different energy. The spin degeneracy of electrons is lifted by Zeeman splitting.

A photon does not interact with magnetic fields, but it also acquires a phase change after a closed loop. By carefully tuning the propagation and coupling phases, Hafezi et al. designed~\cite{hafezi2011robust} a lattice of optical resonators in which the photon acquires the same phase as the A-B phase of electrons moving in a uniform magnetic field. Different from a true quantum Hall topological phase, $T$ is not broken in their static and reciprocal resonator array. So time-reversed channels always exist at the same frequencies allowing back-reflections.
Nevertheless, the authors were able to observe the edge states at near-infrared wavelength~(1.55$\mu m$) in the first set of experiments performed on silicon-on-insulator platform~\cite{hafezi2013imaging}, and in a recent experiment~\cite{mittal2014topologically} also showed that certain robustness against particular types of disorder can still be achieved due to the topological features of the phase arrangements.
%Nevertheless, in the first set of experiments~\cite{hafezi2013imaging,mittal2014topologically} performed on silicon-on-insulator platform, the authors observed the edge states and showed that certain robustness against particular types of disorder can still be achieved due to the topological features of the phase arrangements.

As shown in Fig. \ref{Fig:3_Crows}b, a 2D array of whispering-gallery resonators are spatially coupled through waveguides in between.
Every resonator has two whispering-gallery modes propagating clockwise~(green) and counter-clockwise~(red). They are time-reversed pairs and are similar to the ``spin-up'' and ``spin-down'' degrees of freedom for electrons.
The lengths of the coupling waveguides are carefully designed so that the total coupling phases between resonators precisely match the A-B phase in Fig. \ref{Fig:3_Crows}a: the vertical couplings have no phase changes, while the horizontal couplings have phases linear in $y$.
In each ``spin'' space, photons of opposite circulations experience opposite ``A-B'' phases ($\pm\phi$), just like the electrons in Fig. \ref{Fig:3_Crows}a. These opposite loops are also illustrated in solid and dashed red lines for the ``spin-up'' photons in Fig. \ref{Fig:3_Crows}b.
As a result, the photonic frequency spectrum in this resonator array~\cite{umucalilar2011artificial,petrescu2012anomalous} exhibits both Landau levels and fractal patterns known as the Hofstadter butterfly, which are the signatures of a 2D electron in a uniform magnetic field: the integer quantum Hall effect.
However, without $T$ breaking, the two copies of ``spin'' spaces are degenerate in frequency and couple to each other.
Only under the assumption that these two ``spins'' completely decouple from each other, Chern numbers of same magnitude but opposite signs can be defined and potentially measured~\cite{hafezi2014measuring} in each ``spin'' space. (The ``spin''-polarized counter-propagating edge modes bear similarities to the edge currents in the quantum spin Hall effect for electrons, but they are fundamentally different in symmetry protections and topological invariants as discussed in Box. 2.)
These photonic gapless edge modes are robust against disorder that does not induce ``spin'' flips.
For example, when a defect edge resonator has a different size and resonance frequency from the bulk resonators, then the edge mode will find another route to pass around this defect resonator.
A recent study~\cite{liang2013optical,pasek2013network} suggested that the required spatially-varying couplers along $y$ can be made identical and periodic, achieving otherwise the same phenomena.
Unfortunately, in these reciprocal schemes, perturbations inducing ``spin'' flips are practically ubiquitous: local fabrication imperfections on the resonators or the couplers and even the coupling processes themselves can mix the ``spins'' and induce back-scattering.

This back-scattering in the above time-reversal-invariant systems can be eliminated by breaking $T$, for example using  spatially-coherent time-domain modulations, as proposed theoretically in Ref.~\cite{fang2012realizing}. Fang et al. placed two kinds of single-mode resonators in the lattice shown in Fig. \ref{Fig:3_Crows}c. When the nearest-neighbor coupling is dominant, the two resonators (having different resonance frequencies)
can only couple through the time-harmonic modulation between them. The vertical coupling phases are zero and the horizontal coupling phases increase linearly along the $y$ coordinate, producing effective A-B phases from a uniform magnetic field~\cite{fang2012photonic,li2014photonic,tzuang2014nonreciprocal}.
Photons moving in opposite directions have opposite phases, so they have different frequencies.
Floquet's theorem in the time domain --- similar to the Bloch's theorem in the spatial domain--- is used to solve this lattice system of time-periodic modulations. The resulting Floquet bandstructure has the same gapless edge states as that of a static quantum Hall phase.

Achieving accurate and coherent time-harmonic modulations of a large number of resonators is challenging towards optical frequencies. Rechtsman et al.~\cite{rechtsman2013photonic} translated the modulation from the time domain to the spatial domain, leading to the experimental demonstration of the photonic analogue of the quantum Hall effect at optical frequencies (633nm). These are also the first experiments on Floquet topological phases~\cite{kitagawa2010topological,lindner2011floquet}.
Starting with a 2D resonator array, the authors extended the cavities along the third direction ($z$), obtaining a periodic array of coupled waveguides propagating along $z$.
In their system, $z$ plays the role of time. 
More specifically, the paraxial approximation of the Maxwell's equations results in an equation governing diffraction~(propagating in $z$) that is equivalent to the Schr\"{o}dinger's equation evolving in time.
The periodic helical modulations~\cite{kopp2004chiral,jia2009nonlinear} in $z$ break $z$-symmetry, which is equivalent to the time-domain modulations that break $T$. This symmetry-breaking  opens up protected band degeneracies in the Floquet bandstructure, forming a topologically non-trivial bandgap that contains protected gapless edge modes.

We note that the idea of creating effective magnetic fields for neutral particles~\cite{dalibard2011colloquium} using synthetic gauge fields were first studied in optical lattices~\cite{jaksch2003creation}. Very recently, similar gauge fields were also studied in \emph{optomechanics}~\cite{Schmidt2013Optomechanical} and radio-frequency circuits~\cite{jia2013time}.
Finally,  although approximations like nearest-neighbor in space or rotating-wave in time were adopted in the analysis of the systems described in this section, these higher order corrections do not fundamentally alter the topological invariants and phenomena demonstrated.

\section{Bianisotropic metamaterials}
In bi-anisotropic materials ($\chi\neq 0$ in Eq. \ref{eq:Maxwell} in Box 2)~\cite{kong1972theorems}, the coupling between electric and magnetic fields provides a wider parameter space for the realization of different topological phases.
In particular, it has been shown that bi-anisotropic photonic crystals can achieve topological phases without breaking $T$ ($T$-invariant), so neither magnetism nor time-domain modulations are needed for the topological protection of edge states.
Bi-anisotropic responses are known as optical activity in chiral molecules in nature and can be designed in metamaterials as well.

Bianisotropy acts on photons in a similar way as  spin-orbit coupling does on electrons~\cite{jonsson2006photospin}. In their inspiring theoretical proposal~\cite{khanikaev2013photonic}, Khanikaev et al. enforced polarization (``spin'') degeneracy for photons by equating $\epsilon$ to $\mu$ ($\epsilon=\mu$), so that the transverse electric~(TE) and TM modes in 2D are exactly degenerate in frequencies. When the pseudo-tensor $\chi$ is of the same form as the gyroelectric or gyromagnetic terms in $\epsilon$ or $\mu$, then $\chi$ acts as a magnetic field on each polarization with opposite signs without breaking $T$. This system can be separated into two independent ``spin'' subspaces, in which quantum anomalous Hall phases exist with opposite Chern numbers.

Very recently, it was suggested that the $\epsilon=\mu$ condition could potentially be relaxed~\cite{ma2014topologically}.
Indeed in their experimental work, Chen et al.~\cite{chen2014experimental} relaxed the material requirements to matching the ratio of $\epsilon/\mu$. They also realized broadband effective bianisotropic response by embedding the $\epsilon/\mu$-matched materials in a metallic planar waveguide. These advances enabled them to experimentally observe the % spin-polarized robust 
edge transport around 3GHz.

Similar to the $T$-invariant resonator arrays in Fig. \ref{Fig:3_Crows}b and Ref. ~\cite{hafezi2011robust,hafezi2013imaging,mittal2014topologically,liang2013optical,pasek2013network}, the above metamaterial realizations also require strict conditions in order to decouple the two copies of ``spins''. In these cases, the requirements are on the accurate realization of the constitutive parameters during the metamaterial manufacturing. The lack of intrinsic $T$-protected quantum spin Hall topological phase is one of the most fundamental differences between electronic and photonic systems as discussed in Box 2.

Finally, gapless surface states were also reported to exist in a bulk \emph{hyperbolic metamaterial} having bi-anisotropic responses~\cite{gao2014chiral}.

\section{Quasicrystals}
Quasicrystals are aperiodic structures possessing spatial order. They also have frequency gaps and interfacial states. Quasicrystals can be constructed from the projections of periodic crystals of higher dimensions.

Krauss et al.~\cite{kraus2012topological} projected the 2D quantum Hall phase to a 1D quasicrystal model containing a tunable parameter that is equivalent to the wavevector in 2D. Scanning this periodic parameter reproduces the full gapless frequency spectrum of the 2D quantum Hall phase; that is, the edge-mode frequency of the 1D quasicrystal continuously sweeps through the bulk gap. Experimentally, 1D optical waveguide arrays were fabricated to be spatially varying along the propagation direction $z$ according to the continuous tuning of this parameter. In their system, $z$ plays the role of time. The edge state was observed, starting from one edge of the waveguide array, merging into the bulk modes, then switching to the other edge of the array.
Therefore, light is adiabatically transferred in space from edge to edge. Going a step further, they proposed to realize the quantum Hall phase in 4D using 2D quasicrystals~\cite{kraus2013four}.

\section{Weyl points and line nodes: towards 3D topological phases}
2D Dirac points are the key bandstructures that led to the first proposal and experiments of the photonic analogue of the quantum Hall effect.
For 3D~\cite{chen2011observation,lu2012waveguiding} topological phases, the key bandstructures are line nodes~\cite{Burkov2011nodal}, 3D Dirac points~\cite{Young20123D-Dirac} and, more fundamentally, the Weyl points~\cite{Wan2011Weyl}. However, Weyl points have not been realized in nature. Recently, Lu et al. theoretically proposed~\cite{lu2013weyl} to achieve both line nodes and Weyl points in gyroid photonic crystals realizable at infrared wavelengths using germanium or high-index glasses.

\begin{figure}[h!]
\includegraphics[width=0.46\textwidth]{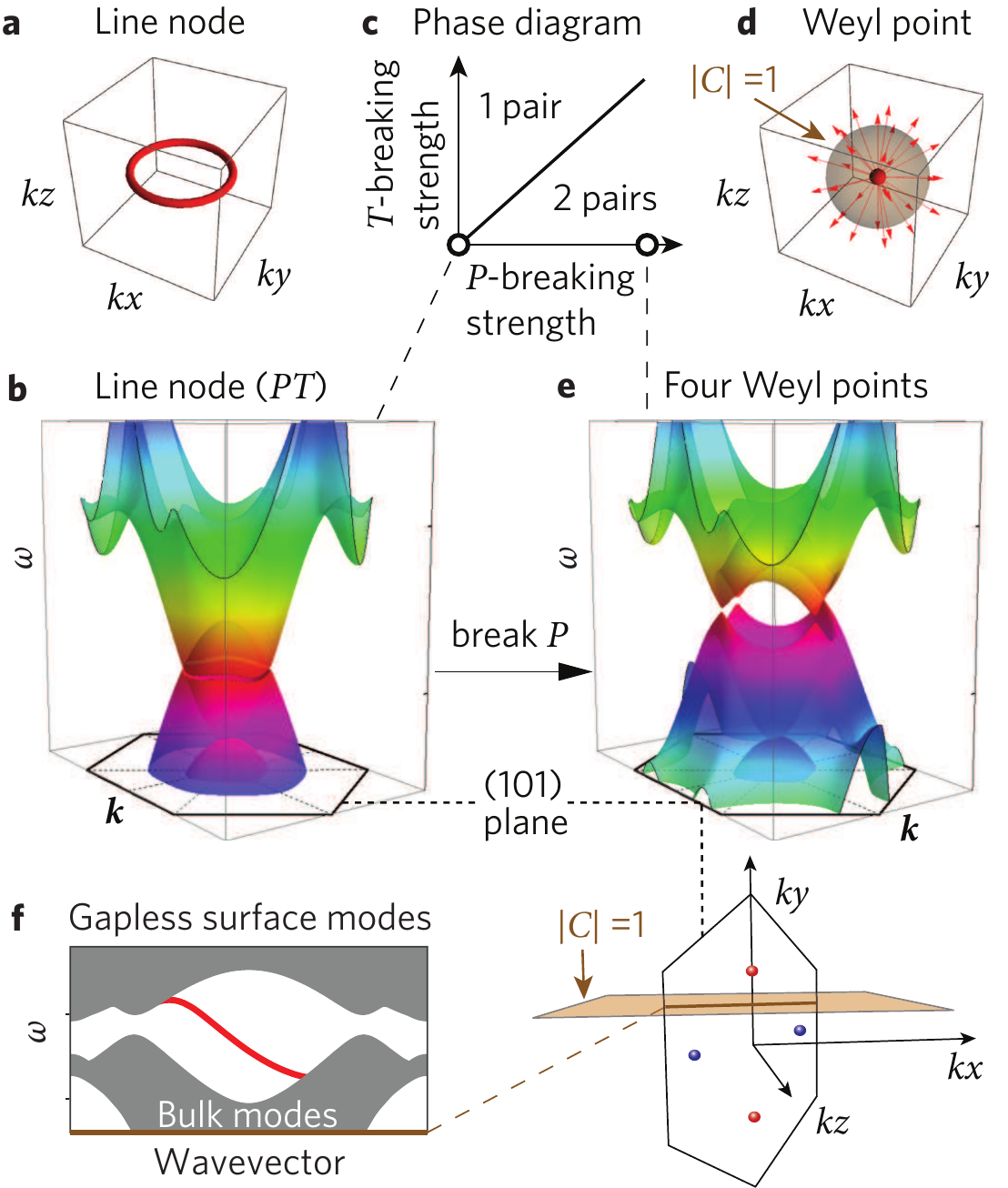}
\caption{\textbf{Phase diagram of line nodes and Weyl points in gyroid photonic crystals.}
\textbf{a}, An illustration of the iso-frequency (red) line of a line node in the 3D momentum space.
\textbf{b}, The line-node bandstructure of a particular $PT$-symmetric DG photonic crystal.
\textbf{c}, The phase diagram of 3D linear degeneracies under $T$ and $P$ breaking perturbations. One pair and two pairs of Weyl points are the minimum numbers in the $T$-dominated and $P$-dominated phases. Line nodes are protected when both symmetry-breaking strengths are zero.
\textbf{d}, An illustration of a Weyl point carrying Chern number of one. A Weyl point is the source or drain of Berry curvatures (red arrows).
\textbf{e}, The Weyl-point bandstructure of a $P$-broken DG photonic crystal.
\textbf{f}, Right side: a plane in the 3D Brillouin zone enclosing unpaired Weyl point has non-zero Chern numbers. The red and blue colors indicate the opposite Chern numbers of the Weyl points. Left side: the surface states on this plane have protected gapless dispersions, plotted along the same wavevector line of brown color on the right.
}
\label{Fig:4_Weyl}
\end{figure}

A line node~\cite{Burkov2011nodal} is a linear line-degeneracy; two bands touch at a closed loop (Fig. \ref{Fig:4_Weyl}a) while being linearly dispersed in the other two directions; it is the extension of Dirac cone dispersions into 3D. For example,  $H(\mathbf{k})=v_xk_x\sigma_x+v_yk_y\sigma_z$ describes a line node along $k_z$. 
So $PT$ protects both Dirac cones and line nodes.
The line node bandstructure in Fig. \ref{Fig:4_Weyl}b is found in a double gyroid (DG) photonic crystal with both $P$ and $T$. The surface dispersions of a line-node photonic crystal can be flat bands in controlled areas of the 2D surface Brillouin zone. When $PT$ is broken, a line node can either open up a gap or split into Weyl points. A phase diagram of the DG photonic crystals is shown in Fig. \ref{Fig:4_Weyl}c, where the line node splits into one pair and two pairs of Weyl points under $T$ and $P$ breakings.

A Weyl point~\cite{Wan2011Weyl} is a linear point-degeneracy; two bands touch at a single point (Fig. \ref{Fig:4_Weyl}d) while being linearly dispersed in all three directions.
The low frequency Hamiltonian of a Weyl point is $H(\mathbf{k})=v_xk_x\sigma_x+v_yk_y\sigma_y+v_zk_z\sigma_z$. Diagonalization leads to the solution $\omega(\mathbf{k})=\pm\sqrt{v_x^2k_x^2+v_y^2k_y^2+v_z^2k_z^2}$.
Since all three Pauli matrices are used up in the Hamiltonian, the solution cannot have a frequency gap.
The existence of the imaginary $\sigma_y$ term means that breaking $PT$ is the necessary condition for obtaining Weyl points.
Illustrated in Fig. \ref{Fig:4_Weyl}d, Weyl points are monopoles of Berry flux; a closed surface in 3D Brillouin zone containing a single Weyl point has a non-zero Chern number of value $\pm 1$. This means a single Weyl point is absolutely robust in the 3D momentum space;
Weyl points must be generated and annihilated pairwise with opposite Chern numbers. 
Since $T$ maps a Weyl point at $\mathbf{k}$ to $\mathbf{-k}$ without changing its Chern number, when only $P$ is broken the minimum number of pairs of Weyl points is two. When only $T$ is broken ($P$ preserved), the minimum number of pairs of Weyl points is one. The bandstructure in Fig. \ref{Fig:4_Weyl}e containing the minimum of four Weyl points is realized in a double gyroid photonic crystal under $P$-breaking. We note that a Dirac point in 3D~\cite{Young20123D-Dirac} is a linear point-degeneracy between four bands, consisting of two Weyl points of opposite Chern numbers sitting on top of each other in frequency.

A photonic crystal containing frequency isolated Weyl points has gapless surface states. Consider the brown plane in the bulk Brillouin zone as shown on the right of Fig. \ref{Fig:4_Weyl}f, it encloses the top red Weyl point ($C=+1$) or equivalently the lower three Weyl points depending on the choice of direction. Either way, this plane has a non-zero Chern number similar to the 2D Brillouin zone in the quantum Hall case. So surface states of that fixed $k_y$ are also gapless and unidirectional. Plotted on the left of Fig. \ref{Fig:4_Weyl}f is an example of these surface states of the P-broken DG photonic crystal.

\section{Outlook}
During the past few years, topological photonics has grown exponentially. Non-trivial topological effects have been proposed and realized in a variety of photonic systems at different wavelengths in all three spatial dimensions. In this review we introduced the main topological concepts, experiments, and proposals, focusing on 2D and 3D realizations. 1D examples are discussed in Ref. \cite{malkova2009observation,tan2014photonic,poshakinskiy2014radiative,xiao2014surface,poddubny2014topological,ling2014plasmonic,poli2014selective}. 

In the coming years, we expect the discovery of new topological mirrors, phases and invariants that could be classified with respect to different symmetries~\cite{Kitaev2009,Schnyder2008,Fu2011TCI,yannopapas2011gapless,denittis2014on}.
The topological phases of interacting photons~\cite{chen2012symmetry,umucalilar2012fractional,hafezi2013non} could be explored by considering  nonlinearity~\cite{lumer2013self} and entanglement.
Various topologically-protected interfacial states between different topological mirrors will be studied.
The immunity to disorder and Anderson localization of those interfacial states need to be addressed. Moreover, the concepts and realizations of topological photonics can be translated to other bosonic systems like \emph{surface plasmons}, excitons~\cite{yuen2014topologically}, exciton-polaritons~\cite{jacqmin2014direct,karzig2014topological}, phonons~\cite{prodan2009topological,kane2013topological} and magnons~\cite{shindou2013topological}. Certain other known robust wave phenomena can be explained through topological interpretations~\cite{zhen2014topological}.

Technologically, the exploitation of topological effects could dramatically improve the robustness of photonic devices in the presence of imperfections. As a result, robust devices are easier to be designed. For example, designers can worry much less about insertion loss and Fabry-Perot noise due to back-reflections.
Topologically-protected transport could solve the key limitation from disorder and localization in slow light~\cite{yang2013experimental} and in coupled resonator optical waveguides~\cite{mittal2014topologically}.
Unidirectional waveguides could decrease the power requirement for classical signals and improve the coherence in quantum links~\cite{koch2010time,aspuru2012observation}.
One-way edge states of $T$-breaking topological phases could be used as compact optical isolators~\cite{jalas2013what}.
Edge states of $T$-invariant topological phases~\cite{khanikaev2013photonic,chen2014experimental} do not have reflection even when the system is reciprocal, thus it might be possible that isolators are unnecessary for photonic circuits consisting of $T$-invariant topological phases.
The realizations of practical topologically-protected unidirectional waveguides towards optical frequencies are currently the main challenge of this emerging field.

Much like the field of topological insulators in electronics, topological photonics promises an enormous variety of breakthroughs in both the fundamental physics and the technological outcomes.

\section{Acknowledgements}
L.L. would like to thank Liang Fu, Chong Wang, Alexander Khanikaev for discussions.
We thank Paola Rebusco and Chia Wei Hsu for critical reading and editing of the manuscript.
J.J. was supported in part by the U.S.A.R.O. through the ISN, under Contract No.W911NF-07-D-0004.
L.L. was supported in part by the MRSEC Program of the NSF under Award No. DMR-0819762. 
M.S. and L.L. were supported in part by the MIT S3TEC EFRC of DOE under Grant No. DE-SC0001299.

\bibliography{Ling}

\onecolumngrid
\clearpage
\setcounter{table}{0}\renewcommand{\thetable}{B\arabic{table}}
\setcounter{figure}{0}\renewcommand{\thefigure}{B\arabic{figure}}

\section{Box 1 | Topological invariant}
\fancypage{\setlength{\fboxsep}{10pt}\fbox}{}
A closed surface can be smoothly deformed into various geometries without cutting and pasting. 
The Gauss-Bonnet theorem~\cite{nakahara2003geometry} of Eq. \ref{eq:Bonnet}, connecting geometry to topology, states that the total Gaussian curvatures ($\mathcal{K}$) of a 2D closed surface is always an integer. This topological invariant, named genus ($g$), characterizes the topology of the surface: the number of holes within. Examples of surfaces of different geni are shown in Fig. \ref{Fig:0_intro}a.
\begin{eqnarray}
\frac{1}{2\pi}\int_{\textnormal{surface}}\mathcal{K}\mathrm{d}A=2(1-g)
\label{eq:Bonnet}
\end{eqnarray}

\begin{figure*}[h]
\includegraphics[width=1\textwidth]{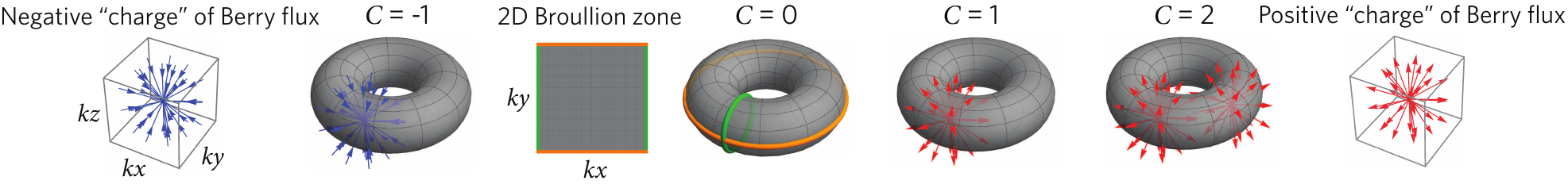}
\caption{\textbf{Chern number as the number of Berry monopoles in momentum space.} A 2D Brillouin zone is topologically equivalent to a torus. The Chern number ($C$) can be viewed as the number of monopoles (charges) of Berry flux inside a closed 2D surface. The arrows represent Berry curvature from the positive and negative charges. In a 3D Brillouin zone, these monopoles are Weyl points.}
\label{Fig:B_ChernNumber}
\end{figure*}

Illustrated in Fig. \ref{Fig:B_ChernNumber}, a two dimensional Brillouin zone is also a closed surface of the same topology of a torus due to the periodic boundary conditions. 
Table B1\ref{tab:Berry} lists the definitions~\cite{Xiao2010Berry} of Berry curvature and Berry flux with respect to Bloch wavefunctions in the Brillouin zone by comparing them to the familiar case of magnetic field and magnetic flux in the real space.
Integrating  the Berry curvature over the torus surface yields the topological invariant called ``Chern number'' that measures the total quantized Berry flux of the 2D surface. The Chern number can be viewed as the number of monopoles of Berry flux inside a closed surface, as illustrated in Fig. \ref{Fig:B_ChernNumber}. An efficient way to calculate Chern numbers in discretized Brillouin zones is described in Ref. ~\cite{fukui2005chern}.

Topological invariants can be arbitrary integers ($\mathbb{Z}$) or binary numbers ($\mathbb{Z}_2$, meaning $\mathbb{Z}$ mod 2).
Chern numbers are integers ($C\in\mathbb{Z}$) and the sum of the Chern numbers over all bands of a given system vanishes.

We note that the geometric phase was first discovered in optics by Pancharatnam~\cite{pancharatnam1956generalized} prior to the discovery of the Berry phase~\cite{berry1984quantal}. The first experiments demonstrating the Berry phase were done in optical fibres~\cite{tomita1986observation}.

\begin{table*}[h]
\begin{tabular}{rrc||cll}
vector potential&$\mathbf{A(r)}$&&&$\mathbf{\mathcal{A}(k)}=\langle u(\mathbf{k})|i\nabla_\mathbf{k}|u(\mathbf{k})\rangle$&Berry connection\\
Aharonov-Bohm phase&$\oint_{}\mathbf{A(r)}\cdot\mathrm{d}\mathbf{l}$&&&$\oint_{}\mathcal{A}(\mathbf{k})\cdot\mathrm{d}\mathbf{l}$&Berry phase\\
magnetic field&$\mathbf{B(r)}=\mathbf{\nabla_{r}\times A(r)}$&&&$\mathbf{\mathcal{F}(k)=\nabla_{k}\times\mathcal{A}(k)}$&Berry curvature\\
magnetic flux&$\iint_{}\mathbf{B(r)}\cdot\mathrm{d}\mathbf{s}$&&&$\iint_{}\mathcal{F}\mathbf{(k)}\cdot\mathrm{d}\mathbf{s}$&Berry flux\\
magnetic monopoles&$\#=\frac{e}{h}\oiint_{}\mathbf{B(r)}\cdot\mathrm{d}\mathbf{s}$&&&$C=\frac{1}{2\pi}\oiint_{}\mathcal{F}\mathbf{(k)}\cdot\mathrm{d}\mathbf{s}$&Chern number\\
\end{tabular}
\label{tab:Berry}
\caption{\textbf{Comparison of Berry phase of Bloch wavefunctions and the Aharonov-Bohm (A-B) phase.} The Berry connection measures the local change in phase of the wavefunctions in the momentum space, where $i\nabla_\mathbf{k}$ is a Hermitian operator. Similar to the vector potential and A-B phase, Berry connection and Berry phase are gauge dependent~[$u(\mathbf{k})\rightarrow{e}^{i\phi(\mathbf{k})}u(\mathbf{k})$]. \
The rest of the quantities are gauge invariant.
The Berry phase is  defined only up to multiples of $2\pi$.
The phase and flux can be connected through Stokes' theorem. Here $u(\mathbf{k})$ is the spatially periodic part of the Bloch function; the inner product of $\langle\rangle$ is done in real space. The one-dimensional Berry phase is also known as the Zak phase.}
\end{table*}

\clearpage
\section{Box 2 | Time reversal symmetry}
Symmetry considerations are crucial to the determination of the possible topological phases of a system. For example, the quantum Hall phase requires the breaking of time-reversal symmetry ($T$). On the other hand, in the recently discovered 2D and 3D topological insulators in electronics, $T$-symmetry is required to protect these topological phases characterized by  $\mathbb{Z}_2$ topological invariants.
For example, the 2D topological insulator, also known as the \emph{quantum spin Hall effect}~\cite{KaneMele2005b}, allows the coexistence of counter-propagating spin-polarized gapless edge states. Without $T$-symmetry however, these edge states can scatter into each other. The edge energy spectrum opens a gap and the insulator can continuously connect to trivial insulators like vacuum.
A large table of symmetry-protected topological phases have been theoretically classified~\cite{Kitaev2009,Schnyder2008}.
These systems have robust interfacial states that are topologically protected only when the corresponding symmetries are present~\cite{Fu2011TCI}.

Here we point out the fundamental difference in time-reversal symmetry between electrons and photons.
A photon is a neutral non-conserved noninteracting spin-1 Boson satisfying the Maxwell's equations, while an electron is a charged conserved interacting spin-$\frac{1}{2}$ Fermion satisfying the Schr\"{o}dinger's equation.
Similar to the Schr\"{o}dinger's equation, the lossless Maxwell's equations at non-zero frequencies can be written as a generalized Hermitian eigenvalue problem in Eq. \ref{eq:Maxwell}.
\begin{eqnarray}
i\left(\begin{array}{cc}
0&\nabla\times\\
-\nabla\times&0
\end{array}\right)
\left(\begin{array}{c}
\mathbf{E}\\\mathbf{H}
\end{array}\right)
=\omega
\left(\begin{array}{cc}
\epsilon&\chi\\
\chi^\dagger&\mu
\end{array}\right)
\left(\begin{array}{c}
\mathbf{E}\\\mathbf{H}
\end{array}\right),
\label{eq:Maxwell}
\end{eqnarray}
where $\epsilon^{\dagger} = \epsilon$, $\mu^{\dagger} = \mu$ and $\chi$ is the bianisotropy term, where $\dagger$ is Hermitian conjugation.

The anti-unitary time $T$ operator is 
$\left(\begin{array}{cc}
1&0\\
0&-1
\end{array}\right)K$, that squares to unity, where $K$ is complex conjugation ($^*$). When $\epsilon^{*} = \epsilon$, $\mu^{*} = \mu$ and $\chi=-\chi^{*}$, the system is $T$-invariant.
Rotating the spin by $2\pi$ is the same as applying the $T$ operator twice ($T^2$). But $T^2$ has different eigenvalues for photons ($T^2=+1$) and electrons ($T^2=-1$). It is this minus sign that ensures the Kramer's degeneracies for electrons at $T$-invariant $\textbf{k}$ points in the Brillouin zone, providing the possibility of gapless connectivity of edge dispersions in the bulk gap.
This fundamental distinction results in different topological classifications of photons and electrons with respect to $T$. For example, photons do not have the same topological phases of 2D or 3D topological insulators protected by $T$ for electrons.

\end{document}